\documentstyle[aps,pre,psfig]{revtex}
\begin{document}

\twocolumn[\hsize\textwidth\columnwidth\hsize\csname
@twocolumnfalse\endcsname
\title{Data clustering and noise undressing of correlation matrices}
\author{Lorenzo Giada and Matteo Marsili}
\address{Istituto Nazionale per la Fisica della Materia (INFM),
Trieste Unit, \\and International School for Advanced Studies (SISSA),
V. Beirut 2-4, Trieste I-34014}
\date{\today}
\maketitle
\widetext

\begin{abstract}
We discuss a new approach to data clustering. We find that maximum 
likelihood leads naturally to an Hamiltonian of Potts variables which 
depends on the correlation matrix and whose low temperature
behavior describes the correlation structure of the data. 
For random, uncorrelated data sets no correlation structure emerges. On
the other hand for data sets with a built-in cluster structure, the method
is able to detect and recover efficiently that structure. Finally we apply the
method to financial time series, where the low temperature behavior reveals 
a non trivial clustering. 
\end{abstract}

]
\narrowtext

\section{Introduction}
Statistical mechanics typically addresses the question of how
structures and order arising from interactions in extended systems are
dressed, and eventually destroyed, by stochastic -- so-called thermal
-- fluctuations.  The inverse problem, unraveling the structure of
correlations from stochastic fluctuations in large data sets, has only
recently been addressed using ideas of statistical
mechanics\cite{Rose,Domany}. This is the case of data clustering
problems, where the goal is to classify $N$ objects, defined by $D$
dimensional vectors $\{\vec \xi_i\}_{i=1}^N$, in equivalence classes.
The general idea\cite{Rose} consists in postulating a cost function
that measures how a possible data structure compares with the sample
one is studying. This cost function can be considered as an
Hamiltonian whose low energy states correspond to the cluster
structures which are mostly compatible with the data sample.
Structures are identified by configurations ${\cal S}=\{s_i\}_{i=1}^N$
of class indices, where $s_i$ is the equivalence class to which object
$i$ belongs.  Regarding $s_i$ as Potts spins, a Potts Hamiltonian
$H_q=-\sum_{i<j} J_{i,j}\delta_{s_i,s_j}$ has been recently proposed
\cite{Domany} as a cost function, with couplings $J_{i,j}$ decreasing
with the distance $d_{i,j}=||\vec \xi_i-\vec \xi_j||$ between objects
$i$ and $j$. The underlying structure of data sets emerges as the
clustering of Potts variables at low temperatures.

In the present work we address the question of data clustering. Rather
than postulating the form of the Hamiltonian, we start from a
statistical {\em Ansatz} and invoke maximum likelihood and maximum
entropy principles.  In this way, the structure of the Hamiltonian
arises naturally from the statistical {\em Ansatz}, without the need
of assumptions on its form. The method is particularly suited to study
high dimensional data sets, where each object is characterized by a
large number $D\gg 1$ of properties. Time series are an ideal example
of high dimensional objects.
The study of the structure of correlations between time series is
therefore a crucial benchmark for our method.


First we derive the form of the Hamiltonian in the general case.  Then
we study the thermal and the ground state properties of this
Hamiltonian by Monte Carlo methods, in three different cases: {\em 1)}
a synthetic uncorrelated data set {\em 2)} a synthetic data set with a
known correlation structure and finally {\em 3)} a data set composed
of financial time series with unknown correlations. We find that
{\em 1)} for random uncorrelated time series no persistent structure
emerges at low temperatures; {\em 2)} if the time series are generated
with some cluster structure ${\cal S}^\star$, we find a phase
transition to a low temperature phase which is dominated by cluster
configurations close to ${\cal S}^\star$. Hence the method does not
introduce spurious correlations and is able to recover known
correlation structures. The nature of the transition is investigated 
by a simple mean field calculation in a simple case. This reveals that 
the phase transition is of first order. 

The financial time series that we will study consists
of the returns of the assets composing the S\&P500 index, whose
correlations have been the subject of much recent
interest\cite{focus,Mantegna,KKM}. On one side it has been observed
\cite{focus} that the S\&P500 correlation matrix is affected by
considerable noise-dressing. Indeed its spectral properties are close
to those of random, uncorrelated time series.  On the other hand,
these same correlation matrices have revealed a non-trivial structure
of correlations when analyzed by minimal spanning tree methods
\cite{Mantegna} and by the method \cite{KKM} of
ref.~\cite{Domany}. These apparently contradictory results raise the
issue of disentangling in a systematic way the effects of fluctuations
from real correlations in a large but finite data set. This is the
main issue we shall focus here.

Quite interestingly, our analysis of the S\&P500 data set reveals a
low temperature behavior dominated by few clusters of correlated
assets with scale invariant properties. We shall not enter into the
details of the economic meaning of our findings, which shall be
discussed elsewhere \cite{GMV}. Our aim is rather to address the
problem of revealing the structure of {\em bare} correlations hidden
in a finite data set. We show that a thermal average over the relevant
cluster structures provides a good fit of the financial correlations,
which allows us to estimate the {\em noise-undressed} correlation
matrix. Finally, we discuss several generalizations of our approach
to generic data clustering.

\section{The Method}\label{two}

Let the data set $\Xi=\{\vec\xi_i\}_{i=1}^N$ be composed of $N$ sets
$\vec\xi_i=\{\xi_i(d)\}_{d=1}^D$ of $D$ measures. 
These are normalized
to zero mean $\sum_d \xi_i(d)/D=0$ and unit variance $\sum_d
\xi_i^2(d)/D=1$. 
For example, in our application below $\xi_i(d)$ is
the normalized daily returns of asset $i$ of the S\&P500 index, in day
$d$. The data set can also refer to a set of $N$ objects which are
characterized by $D$ measured quantities. In this case $\xi_i(d)$ is
the ``normalized'' value of property $d$ for object $i$.  We assume
that $\xi_i(d)$ are Gaussian variables.  The reason is that we want to
focus exclusively on pairwise correlations and the Gaussian model is
the only one which is completely specified at this level. We shall
discuss later how deviations from Gaussian statistics can be accounted
for.
The key quantity of interest is the matrix
\begin{equation}
C_{i,j}(D)\equiv \frac{1}{D}\sum_{d=1}^D\xi_i(d)\xi_j(d).
\label{corr}
\end{equation}

In order to investigate the structure of correlations, let us assume
that $\xi_i(d)$ were generated by the equation
\begin{equation}
\xi_i(d)=\frac{\sqrt{g_{s_i}}\eta_{s_i}(d)+\epsilon_{i}(d)}
{\sqrt{1+g_{s_i}}}.
\label{ansatz}
\end{equation}
Here $g_s>0$ and $s_i$ are integer variables (so-called Potts spins),
$\eta_s(d)$ and $\epsilon_i(d)$ are {\em iid} Gaussian variables with
zero average and unit variance. 

The {\em Ansatz} of Eq.~(\ref{ansatz}) was proposed by Noh \cite{Noh}
to explain the spectral properties found in ref.~\cite{focus}.  The
idea behind it is that each set $i$ belongs to one cluster $s_i$ and
that sets $i$ and $j$ in the same cluster ($s_i=s_j=s$) are correlated
($C_{i,j}\approx g_s/(1+g_s)$) whereas sets in different clusters
($s_i\neq s_j$) are independent. The $s^{\rm th}$ cluster is composed
of $n_s$ sets with {\em internal correlation} $c_s$, where
\begin{equation}
n_s=\sum_{i=1}^N \delta_{s_i,s},\qquad
c_s=
\sum_{i,j=1}^N
C_{i,j}\delta_{s_i,s}\delta_{s_j,s}.
\label{clsvar}
\end{equation}
In order to allow for totally
uncorrelated sets, we allow $s_i$ to take all integer values up to
$N$.  Hence ${\cal S}=\{s_i\}_{i=1}^N$ describes the structure of
correlations whereas the parameters ${\cal G}\equiv\{g_s\}_{s=1}^N$
tune the strength of these correlations.

Note that this {\em Ansatz} is different from the explicative factor
model used in financial applications \cite{CAPM}, which is
discussed in the Appendix.

The correlation matrix generated by Eq.~(\ref{ansatz})
for $D\to\infty$ is 
\begin{equation}
C_{i,j}=\frac{g_{s_i}\delta_{s_i,s_j}+\delta_{i,j}}{1+g_{s_i}}.
\label{Cijg}
\end{equation}
Its distribution of eigenvalues is simple: 
To each $s$ with $n_s\ge 1$ there correspond one eigenvalue
\[
\lambda_{s,0}=\frac{1+g_s n_s}{1+g_s}
\] 
and $n_s -1$ eigenvalues $\lambda_{s,1}=1/(1+g_s)$. 
Hence, large eigenvalues correspond to groups of many ($n_s\gg 1$)
sets. For $D$ finite, we expect noise to lift degeneracies 
between $\lambda_{s,1}$ but to leave the structure of large eigenvalues 
unchanged.

In order to fit the data set $\Xi$ with Eq.~(\ref{ansatz}), let us 
compute the likelihood. This is the probability $P(\Xi|{\cal S},{\cal G})$
of observing the data $\Xi$ as a realization of Eq.~(\ref{ansatz}) with 
structure ${\cal S}$ and 
parameters ${\cal G}=\{g_s\}_{s=1}^N$, and it reads
\[
P(\Xi|{\cal S},{\cal G})=\prod_{d=1}^D
\left\langle\prod_{i=1}^N
\delta\left(
\xi_i(d)-\frac{\sqrt{g_{s_i}}\eta_{s_i}(d)+\epsilon_{i}(d)}
{\sqrt{1+g_{s_i}}}\right)
\right\rangle
\]
where the average is over all the $\eta$'s and $\epsilon$'s variables
and $\delta(x)$ is Dirac's delta function.  Gaussian integration and 
elementary algebra leads to
\begin{eqnarray}
P(\Xi|{\cal S},{\cal G})
&\propto &e^{-DH\{{\cal S},{\cal G}\}}\label{exp}\\
H\{{\cal S},{\cal G}\}&=&\frac{1}{2}\sum_s \big[(1+g_s)
(n_s-\frac{g_s c_s}{1+ g_s n_s}) \nonumber\\
& & -n_s\ln(1+g_s) +\ln(1+ g_s n_s) \big]\label{Hs}.
\end{eqnarray}

For any given structure ${\cal S}$ and $D\gg 1$, the likelihood
$P(\Xi|{\cal S},{\cal G})$ is maximal for $g_s=\hat g_s$, where
\begin{equation}
\hat g_s=\frac{c_s -n_s}{n_s^2-c_s}
\label{xhat}
\end{equation}
for $n_s>1$ and $\hat g_s=0$ for $n_s\le 1$. Inverting
Eq.~(\ref{xhat}) gives $c_s=(\hat g_s n_s^2+n_s)/(\hat g_s+1)$ which
is exactly what one would get combining Eqs.~(\ref{clsvar},\ref{Cijg}). 
Hence the maximum likelihood estimators $\hat g_s$ are consistent
with our Ansatz (\ref{ansatz}).

Note that for uncorrelated sets $C_{i,j}=\delta_{i,j}$ we have
$c_s=n_s$ for each $s$ and hence $\hat g_s=0$. The coupling strength
$\hat g_s$ instead diverges for totally correlated sets ($C_{i,j}=1$)
because $c_s=n_s^2$.

An expansion to second order in $g_s-\hat g_s$ of Eq.~(\ref{Hs}) shows
that the likelihood quickly vanishes for $|g_s-\hat g_s|\gg
1/\sqrt{D}$. Hence, for $D\gg 1$, we can simplify things considerably
by setting $g_s=\hat g_s$ in Eq.~(\ref{Hs}).  The likelihood of
structure ${\cal S}$ under {\em Ansatz} (\ref{ansatz}) then takes the
form $P(\Xi|{\cal S})\propto e^{-D H_c}$, where
\begin{equation}
H_c\{{\cal S}\}=\frac{1}{2}\sum_{s: n_s>0}
\left[\log
\frac{c_s}{n_s}+(n_s-1)\log\frac{n_s^2-c_s}
{n_s^2-n_s}\right].
\label{Hc}
\end{equation}

The ground state ${\cal S}_0$ of $H_c$ yields the maximum likelihood
fit with Eq.~(\ref{ansatz}). This would probably take the Ansatz (\ref{ansatz})
too seriously. In general, it is preferable to consider probabilistic
solutions $P\{{\cal S}\}$ and, following ref. \cite{Rose}, we invoke
the maximum entropy principle: Among all distributions $P\{{\cal S}\}$
with the same average log-likelihood, we select that which has maximal
entropy. This, as usual, leads to the Gibbs distribution 
$P\{{\cal S}\}\propto e^{-\beta H_c\{{\cal S}\}}$
where the inverse temperature $\beta$ arises as a Lagrange multiplier.

The Hamiltonian $H_c$ depends implicitly on the Potts spins $s_i$ 
through the cluster variables $n_s$ and $c_s$ of Eq.~(\ref{clsvar}). 
Unlike the Potts Hamiltonian $H_q$, the dependence on $\delta_{s_i,s_j}$ is 
non-linear and it is modulated by $C_{i,j}$.
For $s_i\ne s_j$ for all $i\ne j$ we have $n_s=c_s=1$ for all $s$ and
hence $H_c=0$. This state is representative of the
high temperature ($\beta\to 0$) limit. The low temperature physics of
$H_c$ is instead non-trivially related to the correlation matrix
$C_{i,j}$. Note, that the ferromagnetic state
$s_i=1$ for each $i$, which dominates as $\beta\to\infty$ in
clustering methods based on Potts models\cite{Domany}, is in general
not the ground state of $H_c$. Intuitively we expect that, if the
model of Eq.~(\ref{ansatz}) is reasonable, $H_c$ should have a well
defined ground state and low temperature phase 
which is energetically dominated by this state. In these cases, as in 
ref.~\cite{Domany}, we expect a thermal phase transition\cite{notaglass}.

The form of the Hamiltonian $H_c$ clearly depends on the {\em Ansatz}
(\ref{ansatz}). For example if one takes a {\em factor model} for the
correlations one finds a different Hamiltonian which depends on
different variables, as discussed in the Appendix. Also note that the
present model only describes positive correlations. It is
straightforward to generalize this approach to the case where a
sizeable number of matrix elements $C_{i,j}$ are negative and not
small.  The idea is to introduce spin variables $\sigma_i=\pm 1$ for
each set and modify Eq.~(\ref{ansatz}) by multiplying the right hand
side by $\sigma_i$. This leads us to the analysis of a system where
the Potts variables $s_i$ and the spin variables $\sigma_i$ interact.
An account of this method shall be given elsewhere \cite{GMV}.

\section{The data}

We consider three different data sets, i.e. three different matrices
$C_{i,j}$. For all of them we took $N=443$ time series of length
$D=1599$. The results with shorter time series will also be discussed
below.

The first data set refers to $N$ totally uncorrelated time series
of length $D$. This is obtained, for example, from Eq.~(\ref{ansatz})
with $s_i=i$. The second also is obtained from Eq.~(\ref{ansatz}), but
this time with preassigned structure ${\cal S}^\star$ and coupling
strengths ${\cal G}^\star$. We shall discuss below how the specific
structure was chosen.  These two data sets help us to understand how
the method performs when no structure at all is present and to check
whether a predefined structure can be recovered.

Our third data set is made of financial time series of asset prices
relative to the Standard \& Poors 500 index (S\&P500). More precisely
$\xi_i(d)$ is the normalized daily returns of asset $i$ of the S\&P500
index, in day $d$; this is defined as
\begin{equation}\label{returns}
\xi_i(d)=
\frac{\log [p_i(d)/p_i(d-1)]-r_i}{\sigma_i},
\end{equation}
where $p_i(d)$ is the price of asset $i$ in day $d$. The parameters
$r_i$ and $\sigma_i$ are determined in order to have zero mean and
unit variance for all $i$.

Correlation matrices of financial time series are of great practical
interest. Indeed they are at the basis of risk minimization in the
modern portfolio theory\cite{CAPM}. This states that, in order to
reduce risk, the investment needs to be {\em diversified}
(i.e. divided) on many uncorrelated assets, so that price fluctuations
are averaged out.  However the measure of correlation in samples with
a number of observation times comparable to the number of assets was
recently found to be affected by considerable noise-dressing
\cite{focus}. For example the S\&P500 is composed of $N=500$ assets
and considering daily data from July 3rd 1989 to October 27th 1995 one
has $D=1600$ data points for each asset. This data set is then an 
ideal instance of a problem where our method applies. 

\begin{figure}
\centerline{\psfig{file=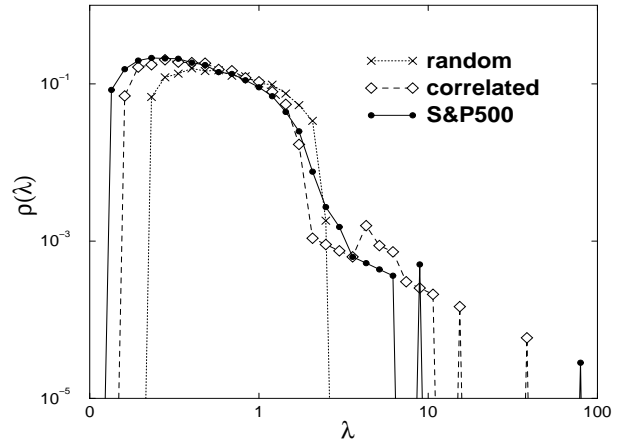,width=8cm,height=6cm}}
\caption{Distribution of eigenvalues of the correlation matrices 
of S\&P500 (full line $\bullet$) random (dotted $\times$) 
and synthetic correlated (dashed $\diamond$) time series.} 
\label{figmata}
\end{figure}

In addition, this data set has been studied by other authors with
several methods including spectral analysis \cite{focus}, minimal
spanning tree \cite{Mantegna} and super-paramagnetic clustering
\cite{KKM}. This allows us to compare the results of our method with
those of other methods. Finally, in order to better appreciate the
performance of our method in a real application, we choose the
synthetic correlated data set $\{{\cal S}^\star,{\cal G}^\star\}$ as a
``large likelihood'' structure of the S\&P500 data set.  In other
words we performed a simulated annealing experiment on the S\&P500
data set, where the fictitious temperature $1/\beta$ was gradually
decreased to $0$. This allowed us to compare how well the real S\&P500
data set can be described by a maximum likelihood
structure\footnote{The maximum likelihood structure may be
computationally hard to find and simulated annealing may get trapped
into a local minimum. Indeed in our case we found slightly different
structures in different experiments. The structure ${\cal S}^\star$
was that with largest likelihood.}.

Figure \ref{figmata} shows a comparison of the spectral properties of
the three correlation matrices. The spectrum of eigenvalues for
uncorrelated time series are known exactly\cite{Sengupta}. It extends
over an interval of size $\sim N/D$ around $\lambda=1$, as shown in
Fig.~\ref{figmata}. The spectrum of eigenvalues of the S\&P500
correlation matrix has a similar shape for $\lambda\approx 1$. This
suggests that significant noise-dressing due to finite $D$
occurs\cite{focus}. The tail of the distribution ($\lambda\gg 1$)
implies that some correlation is however present. Within our
framework, large eigenvalues are associated with large clusters.
Indeed the synthetic correlated data set has a broad distribution of
cluster sizes (see Fig.~\ref{figcls}) and a correspondingly fat tail
in the distribution of eigenvalues.

\section{Clustering by Monte Carlo simulations}

In order to study the properties of $H_c$ we resort to Monte Carlo
(MC) method with Metropolis algorithm\cite{Metropolis}.  This, at
equilibration, allows us to sample the Gibbs distribution $P\{{\cal
S}\}$ and compute average quantities, such as the internal energy
$E_\beta=\langle{H_c}\rangle_\beta$ where
$\langle{\ldots}\rangle_\beta$ stands for thermal average.  To detect
the occurrence of spontaneous magnetization -- which correspond to the
$s_i$ remaining locked into energetically favorable configurations at
low temperature -- we measure the autocorrelation function
\begin{equation}
\chi(t,\tau)=\frac{\sum_{i<j}\delta_{s_i(t),s_j(t)}
\delta_{s_i(t+\tau),s_j(t+\tau)}}{\sum_{i<j}
\delta_{s_i(t),s_j(t)}}.
\label{chit}
\end{equation}
This quantity tells us how many pairs of sites belonging to the same
cluster at time $t$ are still found in the same cluster after $\tau$ MC steps.
For $t$ large enough, $\chi$ becomes a function of $\tau$ only. 
This function decreases rapidly to a plateau value 
$\chi_\beta=\langle{\chi(t,\tau)}\rangle_\beta$ for $t\gg \tau\gg 1$. 
Clearly $\chi_\beta\simeq 0$ implies that no persistent
structure is present whereas, at the other extreme, $\chi_\beta=1$ implies that
all sites are locked in a persistent structure of clusters.

\begin{figure}
\centerline{\psfig{file=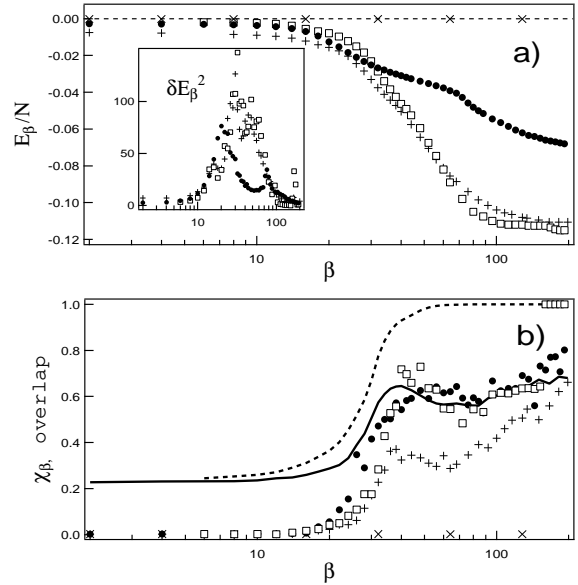,width=8cm}}
\caption{{\bf a)} Energy $E_\beta$ as a function of $\beta$ for random 
($\times$), S\&P500 ($+$) and correlated ($\Box$) data sets of length
$D=1599$ respectively. The results for the S\&P500 data set over the last
$D=400$ days are also shown ($\bullet$). Inset: square energy fluctuation
$\delta E_\beta^2$ vs $\beta$ for the same data sets (same symbols). 
{\bf b)} Autocorrelation $\chi_\beta$ as a function of $\beta$ for the 
same data sets (same symbols). The full (dashed) line refers to the 
overlap with the configuration $s^\star$ for the S\&P500 (correlated)
data set with $D=1599$.}
\label{figeco}
\end{figure}

We monitored these quantities for the three data sets.  Let
us start with a truly uncorrelated time series. We generate the time
series $\xi_i(d)$ and then compute $C_{i,j}$ by Eq.~(\ref{corr}).
With this we compute the Hamiltonian $H_c$ and study its thermal
properties by the MC method. We do not expect any clustering to emerge
in this case. Indeed, the internal energy $E_\beta$ stays very close
to $0$ (see Fig.~\ref{figeco}a) for all values of $\beta$ investigated
up to $\beta=512$.
Correspondingly no persistent cluster arises, i.e. $\chi_\beta\simeq 0$. 

The results change turning to correlated data. 
Let us first discuss the S\&P500 data (for $D=1599$). 
As Fig.~\ref{figeco}a shows, for 
$\beta\approx 20$ the energy $E_\beta$ starts deviating significantly
from zero. For $\beta>20$ persistent clusters are present:
$\chi_\beta$ rapidly raises from zero and it has a maximum at
$\beta\approx 40$ (see Fig.~\ref{figeco}b). The energy fluctuations
reported in the inset shows a broad peak of intensity marking the
onset of an ordered low temperature phase. As $\beta$ increases the
dynamics is significantly slowed down. At $\beta\approx 200$ the
energy reaches a minimal value $E_\beta\simeq -0.11 N$ and does not
decrease significantly increasing $\beta$ at least up to
$\beta=4095$. This energy is smaller than that of the ferromagnetic
state ($E_f=-0.086 N$), with all sets in the same cluster. The system
in this range of temperatures visits only few configurations.

\begin{figure}
\centerline{\psfig{file=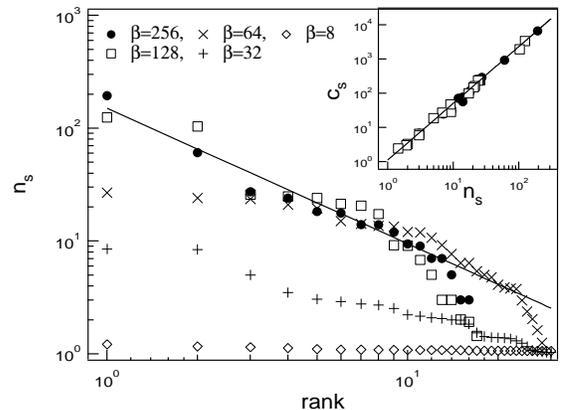,width=8cm}}
\caption{Rank plot of $n_s$ for several values of $\beta$. The line
corresponds to $n\sim {\rm rank}^{-1.2}$.Inset:
$c_s$ versus $n_s$ for $\beta=256$ ($\bullet$) and $\beta$ ($\Box$).
The line corresponds to  $c\sim n^{1.66}$.}
\label{figcls}
\end{figure}

The statistical properties of cluster configurations, as $\beta$
varies, are shown in Fig.~\ref{figcls}. For small $\beta$ only small
clusters survive to thermal fluctuations. As $\beta$ increases a
distribution of cluster sizes develops. At low temperatures the rank
order plot of $n_s$ reveals a broad distribution of clusters with the
largest aggregating more than $190$ sets.  By a power law fit of this
distribution, we find that the number of clusters with more than $n$
sets decays as $n^{-0.83}$. The scatter plot of $c_s$ versus $n_s$
also reveals a non-trivial power law dependence $c_s\sim
n_s^{1.66}$. This gives a statistical characterization of the dominant
configurations of clusters at low energy.
The clusters structure we obtain is reasonable from the economic point
of view: companies in the same economic sector belong to the same
cluster. These issues will be discussed in detail
elsewhere\cite{GMV}. Here we restrict our attention to the
clustering method.

It is instructive to compare these results with those obtained form
shorter time series. We performed a second set of simulations with
$C_{i,j}$ computed using the quotes of the S\&P500 assets for the last
$D=400$ days. We find two inverse temperatures $\beta_1\approx 20$ and
$\beta_2\approx 80$ which separates three regimes. This is signalled
by the bending in the $E_\beta$ curve and by peaks in the $\delta
E_\beta^2$ vs $\beta$ plot. At the first temperature clusters start to
appear. For $\beta<\beta_2$ the largest cluster groups less than $30$
sets and for $\beta>\beta_2$ larger clusters $n_s\approx 100$
appear. This hints at a time dependence of correlations, which are
averaged in the $D=1599$ data set.  For even shorter time series we
found that sampling errors, acting like a temperature, destroy large
clusters and only relatively small clusters ($n_s<40$ for $D=60$) were
found.

Finally let us discuss the results for the synthetic correlated data
set (for $D=1599$). As already mentioned, the structure ${\cal
S}^\star$ is a typical low energy configuration for the S\&P500 data
set extracted from the previous simulations (with $D=1599$). The
parameters $g_s^\star$ where deduced from the $n_s$ and $c_s$ of this
configuration, via Eq.~(\ref{xhat}). We recall that this data set is
useful for two reasons: first it allows one to understand to what
extent a structure of correlation put by hand with the form dictated
by Eq.~(\ref{ansatz}) can be correctly recovered. Secondly it allows
us to compare the results found for the S\&P500 data with those of a
time series with correlations described by Eq.~(\ref{ansatz}).

For $\beta<150$, the behaviors of $E_\beta$, $\delta E_\beta^2$ and
$\chi_\beta$ are similar to those found for the S\&P500 data (see
Fig.~\ref{figeco}). A second, sharp peak in $\delta E_\beta^2$ at
$\beta\approx 170$ signals a new clustering transition. Below this
temperature, as shown by the plot of $\chi_\beta$
(Fig.~\ref{figeco}b), the MC dynamics freezes into the original
structure ${\cal S}^\star$.  The overlap with the configuration ${\cal
S}^\star$, defined as in Eq.~(\ref{chit}) as the fraction of ``bonds''
$s_i=s_j$ for which $s_i^\star=s_j^\star$, quickly converges to $1$
(see Fig.~\ref{figeco}b) for the synthetic time series, whereas it
remains around $60\%$ for the S\&P500 data set.  This, on one hand
means that the original structure ${\cal S}^\star$ can be recovered
quite efficiently. On the other hand, it suggests that several cluster
configuration compete at low temperatures in the S\&P500 data set.

\section{Mean Field model}

In this section we would like to determine the nature of the
clustering transition that takes place in our system.
We apply our method to an unrealistically simple situation that
will allow us to extract analytical expressions for the order
parameter associated with the phase transition. Our analysis is rather
similar to that in \cite{Domany}.

We take $N$ time series that belong to $M$ clusters of the same
size $n=N/M$. Let $\bar s_i$ be the cluster index of the $i^{\rm th}$ time
series 
, and let
\begin{equation}
C_{i,j}=\gamma\delta_{\bar s_i,\bar s_j}+(1-\gamma)\delta_{i,j}
\end{equation}
be the correlation matrix for $D=\infty$. This means that time series
with $\bar s_i=\bar s_j$ have correlation $\gamma$ whereas $\bar
s_i\not =\bar s_j$ have $C_{i,j}=0$. The matrix $C_{i,j}$ has $N/M$
blocks of size $M$ along the diagonal and is zero elsewhere.  A finite
$D$ sample of this problem is generated with Eq.~(\ref{ansatz}) with
$g_s=\gamma/(1-\gamma)$ and $s_i=\bar s_i$.

To fix ideas we can imagine to have the problem of putting $N$ balls
of $M$ different colors in $M$ boxes. Colors represent the original
structure $\bar s_i$ whereas boxes represent the actual clustering
configuration $s_i$. When the balls contained in each box have the
same color, the original cluster structure has been recovered.
Let $m_s^c$ be the number of balls of color $c$ in box (or cluster)
$s$.  Now $\sum_s m_s^c =n=N/M$ is the overall number of balls of
color $c$, assumed equal for all colors, and $\sum_c m_s^c =n_s$ is
the number of balls in box $s$, as in Eq.~(\ref{clsvar}).  With the
above choice of parameters the \emph{internal correlation} of box $s$
for a given configuration $\{ m_s^c \} $ of clusters is
\begin{equation}
\label{cs}
c_s=(1-\gamma ) n_s+\gamma \sum_c {m_s^c}^2.
\end{equation}

To compute the free energy $F=U-TS$ of the system we use the energy
$H_c$ as in Eq.~(\ref{Hc}) and we estimate the configuration entropy in
the following way: The number of ways in which one can distribute the
balls of color $c$ by putting $m_s^c$ of them in box $s$ is
\begin{equation}
\frac{(\sum_s m_s^c)!}{\prod_s (m_s^c !)} = \frac{n!}{\prod_s (m_s^c !)},
\label{config}
\end{equation}
and the total number of configurations for all the colors is the
product over $c$ of this expression. Taking the logarithm of this
product we obtain the configuration entropy
\begin{equation}
S=\log\left(\prod_c\frac{(\sum_s m_s^c)!}{\prod_s (m_s^c !)}
\right)=\sum_{s,c}m_s^c\log\left(\frac{m_s^c}{n}\right),
\end{equation}
where we have approximated in the usual way the logarithm of the
factorial. Finally the free energy is:
\begin{eqnarray}
F&=&\sum_s F_s \nonumber \\
F_s&=&\frac{1}{2}\left[\log\frac{c_s}{n}+(n-1)\log(\frac{n^2-c_s}{n^2-n}) 
\right] +\nonumber \\
& &+T \sum_c m_s^c\log\left( \frac{m_s^c}{n}\right).
\label{fs}
\end{eqnarray}

After substitution of Eq.~(\ref{cs}) we find an expression which
depends on the occupation variables $m_s^c$. The occupation in
different boxes are related by the overall constraints $\sum_{s}
m_s^c=N/M$. We take the mean-field approximation, which is legitimate
in cases like this, where we neglect these effects.  In other words,
we minimize each of the $F_s$ independently and we suppress
therefore the subscript $s$ from now on. 

We can then focus on just one box and look for solutions of the form
\begin{equation}
m^c=\frac{N}{M}
\cases{
\phi &  $c=1$ \cr
\frac{1-\phi}{M-1} & $c\neq 1$},
\end{equation}
with $0\leq \phi\leq1$. In this {\em Ansatz}, the balls of color $c=1$ are
more (or less) numerous that those of other colors $c>1$. In the
spirit of mean-field approximation, we neglect the possibility that
the number of balls of colors $c>1$ may be unevenly distributed. Hence
the parameter $\phi$ plays the role of the order parameter.

The paramagnetic solution $\phi=1/M$, which corresponds to an uniform
distribution of colors, is always a solution of the saddle point
equations $\frac{\partial F}{\partial \phi}=0$.  This state is
expected to be stable (the minimum of $F$) at high temperature.  A
second solution of $\frac{\partial F}{\partial \phi}=0$, which
corresponds to the clustered ``ferromagnetic'' state, appears at
intermediate temperatures with $\phi\approx 1$. For $T=T_c$ the values
of the free energy corresponding to the two states are equal and a
first order phase transition to a ferromagnetic state takes place.
The order of the transition is independent of the values of the
parameters, while the critical temperature is determined mainly by the
number of time series $N$.

\begin{figure}
\centerline{\psfig{file=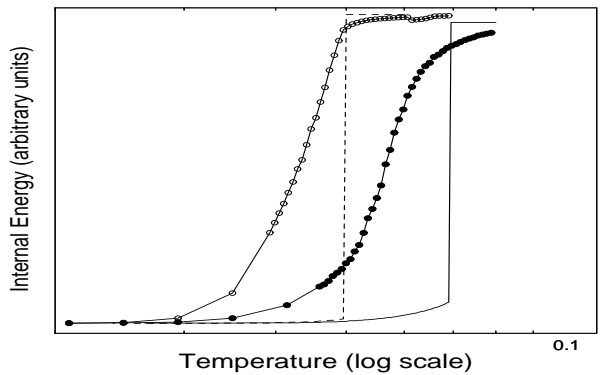,width=8cm,height=5cm}}
\caption{Ferromagnetic clustering: filled dots and full lines
correspond to Monte Carlo and analytical results for a system with
$N=150$, $M=6$; empty dots and dashed lines to $N=2400$,
$M=24$. $\gamma$ is always $0.3$. } 
\label{pic}
\end{figure}

We checked that this result is compatible with that obtained from
Monte Carlo simulations. We find that the mean-field approach provides
a good qualitative picture of the transition and a reliable estimate
of the critical temperature at which it takes place
(see Fig.~\ref{pic}).

\section{Noise Undressing}

Eq.~(\ref{ansatz}) with a single cluster configuration
($\beta\to\infty$), is inadequate to capture the full complexity of
the correlations in the S\&P500 data set.  Probabilistic clustering,
where several cluster structures ${\cal S}$ are allowed with their
Gibbs probability $P\{{\cal S}\}$, provides an alternative
approximation. In this approach the parameter $\beta$ can be tuned to
determine the optimal spread in configuration space, which best
describes the correlation structure built in the original data
set. This line of reasoning will lead us to a method to ``fit'' the
correlation structure of a data set with a single parameter $\beta$.
This will finally allow us to undress the correlation matrix
$C_{i,j}(D)$ of its noise-dressing and to reveal the bare
correlations.

Let us start by remarking that the problem with Eq.~(\ref{ansatz}) is
that it stipulates that a set $i$ can belong to only one cluster. 
This suggests to consider the generalized model

\begin{equation}
\xi_i(d)=\frac{\sum_s\sqrt{g_{s,i}}\eta_s(d)+\epsilon_i(d)}
{\sqrt{1+\sum_s g_{s,i}}},
\label{fitbeta}
\end{equation}
where each set $i$ can belong to any cluster $s$.  Eq.~(\ref{fitbeta})
has, on the other side, the disadvantage that it depends on too many
variables, and it leads to overfitting stochastic fluctuations.

The finite temperature distribution $P\{{\cal S}\}$ provides a natural
way out of this situation. Indeed at finite $\beta$ each set $i$ visits
different clusters $s$ and we can define
\begin{equation}
g_{s,i}(\beta)=\left\langle\frac{c_s-n_s}{n^2_s-c_s}
\delta_{s,s_i}\right\rangle_\beta.
\label{ext_ansatz}
\end{equation}

The parameters $g_{s,i}(\beta)$ can be measured in a MC simulation
and provide us with a measure of the strength of the correlation 
between set $i$ and cluster $s$. 

These parameters and Eq.~(\ref{fitbeta}) also allow us to generate
synthetic data sets $\tilde \xi_i(d)$, whose statistical properties
can be compared to those of the original data set. We make this
comparison using the spectral properties of the correlation matrix. In
other words, with Eq.~(\ref{corr}) and $\tilde \xi_i(d)$ we compute a
``$\beta$-synthetic'' correlation matrix; we determine the spectrum of
eigenvalues and compare it to that of the original matrix. The parameter
$\beta$ can be tuned in order to get the best fit. 

\begin{figure}
\centerline{\psfig{file=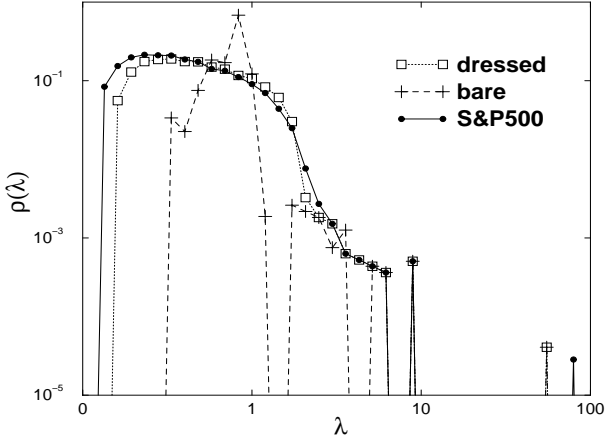,width=8cm,height=6cm}}
\caption{Comparison of the spectrum of the S\&P500 correlation matrix (full line $\bullet$) with noise-dressed (dotted $\Box$) and bare (dashed $+$) correlation matrices generated by Eq.~(\ref{fitbeta}).}
\label{figmatb}
\end{figure}

This procedure was carried out for the S\&P500 data set.  The
eigenvalue spectra of the two matrices are compared in
Fig.~\ref{figmatb} for $\beta=48$.  The value of $\beta$ was chosen by
visual inspection as that giving the best fit. The curves are
remarkably close, suggesting that Eq.~(\ref{fitbeta}) provides a good
statistical description of the correlations among assets.

Once the value $\beta^*$ which gives the best fit is found, we can
compute the {\em noise undressed} correlation matrix
\begin{equation}
C_{i,j}^{*} (\infty)=\frac{\delta_{i,j}+\sum_s\sqrt{g_{s,i}^* g_{s,j}^*}}
{\sqrt{1+\sum_s g_{s,i}^*}\sqrt{1+\sum_s g_{s,j}^*}}
\label{Cijstar}
\end{equation}
from the parameters $g_{s,i}^*=g_{s,i}(\beta^*)$. This is the
correlation matrix of a synthetic data set obtained from
Eq.~(\ref{fitbeta}) with $D\to\infty$.  Fig.~\ref{figmatb} shows the
eigenvalue distribution of the {\em noise undressed} matrix
$C_{i,j}^{*}(\infty)$. This allows one to appreciate the effect
of noise dressing. As expected, noise mainly affects small
eigenvalues.

\section{Discussion and Conclusions}

The applicability of the method can be extended considerably to a
generic data set $\{\vec x_i\}_{i=1}^N$. $\vec x_i$ need not be a time
series. The distribution of $x_i(d)$ need not be Gaussian and it does
not even need to be the same across $i$. For example, $x_i(d)$ may be
the measure of the $d^{\rm th}$ feature of the $i^{\rm th}$ object or
the concentration of species $i$ in the $d^{\rm th}$ sample of an
experiment.  The idea is to map the data set $\vec x_i$ into a
Gaussian time series $\vec\xi_i$ to which we apply Eq.~(\ref{ansatz}).
The mapping results from requiring that non-parametric
cross-correlations $\tau_{i,j}^x=\tau_{i,j}^\xi$ are preserved. To do
this in practice we compute Kendall's $\tau$ \cite{Kendall} for the
$\vec x_i$ data sets: $\tau_{i,j}^{x} =\langle{\hbox{sign}
[x_i(d)-x_i(d')]\hbox{sign} [x_j(d)-x_j(d')]}\rangle_{d<d'}$.  For two
Gaussian time series with correlation $C_{i,j}$ one can compute
analytically $\tau^x_{i,j}$ in the limit $D\to\infty$. This leads to
the relation 
\begin{equation}
C_{i,j}=\frac{\tan(\pi\tau^x_{i,j}/2)}
{\sqrt{1+\tan^2(\pi\tau^x_{i,j}/2)}}
\label{CijTij}
\end{equation}
between Gaussian and non-parametric correlations.  This equation
allows us to translated non-parametric correlations into Gaussian
correlations. From these one can build $H_c$ of Eq.~(\ref{Hc}) and
study the clustering properties.

This procedure has been tested for the S\&P500 data set, for which it
is known that $\xi_i(d)$ has non-Gaussian statistics
\cite{Mantegna&Stanley}. We have found indistinguishable results which
indicate that the deviation from Gaussian behavior have little or no
effect on the results. We expect that this approach breaks down when
the marginal distribution of $\xi_i(d)$ is such that the second moment
is not defined. In that case $C_{i,j}$ computed from $\tau_{i,j}$ and
Eq.~(\ref{CijTij}) can even fail to be positive definite.

With respect to ref.~\cite{Domany}, our approach does not need any
assumption on the form of the Hamiltonian.  As input, the method only
needs the correlation matrix $C_{i,j}$ (or $\tau_{i,j}$). The range of
interactions is set by the correlations themselves. Indeed our method
predicts a non-trivial ground state ${\cal S}_0$ which is not, in 
general, the ferromagnetic one.

For small $D$, the local interaction of ref.~\cite{Domany} may well be
more efficient in capturing the structure of data. Our method is most
useful in cases where $D\sim N\gg 1$. These ideas can clearly be
extended to models of correlations different from Eq.~(\ref{ansatz})
as shown, for example, in the Appendix.

We acknowledge R. Zecchina, R. Pastor-Satorras and D. Vergni for
interesting discussions and R. N. Mantegna for providing the S\&P500
data.
\appendix
\section{}\label{appe}
We define here the {\em explicative factor model} for
stocks returns (also known as {\em  multi index model}, see e.~g.~\cite{CAPM}):
\begin{equation}
\xi_i(t)=\frac{\vec{v}_i\vec{\eta}(t)+\epsilon_i}{\sqrt{1+v_i^2}},\qquad
v_i^2=\vec{v}_i\vec{v}_i=\sum_{\alpha=1}^L (v_i^\alpha)^2
\label{vect}
\end{equation}
where $\vec{v}_i$ are $L$ dimensional vectors and $\vec\eta(t)$ is a
$L$ dimensional Gaussian random vector
$\langle\eta^\alpha(t)\rangle=0$ and
$\langle\eta^\alpha(t)\eta^\beta(t')\rangle=
\delta_{\alpha,\beta}\delta_{t,t'}$.

The idea is that there are $L$ {\em explicative factors}
$\eta^1(t),\ldots, \eta^L(t)$ which describe the fluctuations of each
stock price. This model is different from the one we considered in the
text in that each time series is coupled to all other with a different
strength. This can be easily understood by observing that the model
(\ref{ansatz}) can be cast in the form of an explicative factor model
with $v_i^\alpha=g_{s_i}\delta_{\alpha,s_i}$.  This is a rather
particular form of Eq.~(\ref{vect}). We observe, however, that $L$ must
be much smaller than $N$ in order to avoid problems of overfitting
with Eq.~(\ref{vect}), while Eq.~(\ref{ansatz}) requires $L\approx N$.

As we did in Sect.~\ref{two} we look at the probability of observing
the time series $\xi_i(t)$ given the model Eq.~(\ref{vect}) and the
parameters $\vec{v}_i$:
\begin{equation}
P\{\xi_i(t)|\vec{v}_i\}=\prod_{t=1}^D
\left\langle\prod_{i=1}^N\delta\left(\xi_i(t)-\frac{\vec{v}_i\vec{\eta}+\epsilon_i}
{\sqrt{1+v_i^2}}\right)\right\rangle_{\vec\eta,\epsilon}.
\end{equation}
After taking the average over the Gaussian variables one obtains the
equivalent of Eqs.~(\ref{Hs}) and (\ref{exp})
\[
P\{\xi_i(t)|\vec{v}_i\}  =  e^{-DH\{\vec v\}}
\]
\begin{eqnarray}
H\{\vec v\}&=&
\frac{1}{2}\sum_i\left[(1+v_i^2)-\log(1+v_i^2)\right]-\nonumber\\
&&\frac{1}{2} \hbox{Tr} \log(1+V)-\frac{1}{2}\hbox{Tr}
\frac{\chi}{1+V},\label{Hv} 
\end{eqnarray}
where we have defined the matrices
\begin{eqnarray}
V_{\alpha,\beta}&=&\sum_{i=1}^N v_i^\alpha v_i^\beta\\
\chi_{\alpha,\beta}&=&\sum_{i,j=1}^N
C_{i,j}\sqrt{1+v_i^2}v_i^\alpha\sqrt{1+v_j^2}v_j^\beta,
\end{eqnarray}
and $C_{i,j}$ is defined in Eq.~(\ref{corr}).  We note that the second
term in Eq.~(\ref{Hv}) is sub-extensive, and could be neglected;
nevertheless in the presence of the matrix $\chi$ a Monte Carlo
simulation becomes excessively time consuming, since a change in
$\vec{v}_k$ requires order $N$ operations to compute the new
matrix. This may considerably limit the practical applicability of
this method.


\end{document}